%% file: mainFETA.tex
\documentclass[a4paper]{llncs}
\usepackage{amssymb}
\usepackage{graphicx}
\usepackage{url}
\usepackage[T1]{fontenc}
\usepackage[latin1]{inputenc}
\usepackage{lmodern}
\usepackage{paralist}
\usepackage{tikz}
\usepackage{epstopdf}
\usepackage{xspace}
\usetikzlibrary{calc,chains, positioning,shapes.geometric,arrows,backgrounds,decorations.markings,shadows,calc,shapes.misc}
\usepgflibrary{arrows}
      
\tikzstyle{io} = [trapezium, trapezium left angle=70, trapezium right angle=110, minimum width=3cm, minimum height=1cm, text centered, draw=black] 
\tikzstyle{process} = [rectangle, minimum width=1cm, minimum height=0.5cm, text centered, draw=black] 
\tikzstyle{decision} = [diamond, minimum width=3cm, minimum height=0.5cm, text centered, draw=black] 
\tikzstyle{arrow} = [thick,->,>=stealth]
\tikzstyle{startstop} = [rectangle, rounded corners, minimum width=3cm, minimum height=1cm,text centered, draw=black, fill=red!30]
\tikzset{
              mynode/.style={rectangle,rounded corners,draw=black, top color=white, bottom color=yellow!50,very thick, inner sep=1em, minimum size=3em, text centered},
              myarrow/.style={->, >=latex', shorten >=1pt, thick},
              mylabel/.style={text width=7em, text centered},
         database/.style={
               cylinder,
               cylinder uses custom fill,
               cylinder body fill=blue!20,
               cylinder end fill=blue!20, 
               shape border rotate=90,
               draw,
                minimum height=1.5cm,
                minimum width=1.5cm
             }
          }

      \tikzset{
        sshadow/.style={opacity=.25, shadow xshift=0.05, shadow yshift=-0.06},
      }
      
      \def\schemeld[#1,#2,#3,#4,#5,#6]#7{ %
        \node[draw, diamond, shape aspect=#3, rotate=#2, minimum size=#1, %
        bottom color=gray!55, top color=gray!25, color=gray!65!black, %
        drop shadow={sshadow,color=green!60!black}, #4] (#5) at #6
        {\textcolor{green!40}{}}; %
        \node at #6 {#7};%
      }
      
        \def\schemel[#1,#2,#3,#4,#5,#6]#7{ %
              \node[draw, cylinder, rotate=90, minimum size=#1, %
              bottom color=gray!55, top color=gray!25, color=gray!65!black, %
              drop shadow={sshadow,color=gray!60!black}, #4] (#5) at #6
              {\textcolor{green!40}{}}; %
              \node at #6 {#7};%
            }
      
      \def\schemer[#1,#2,#3,#4,#5,#6]#7{ %
        \node[draw, diamond, shape aspect=#3, rotate=#2, minimum size=#1, %
        bottom color=green!65, top color=green!30, color=green!60!black, %
        drop shadow={sshadow,color=green!65!black}, #4] (#5) at #6
        {\textcolor{green!53}{bla}}; %
        \node at #6 {#7}; %
      }
      
      \def\tboxu[#1,#2,#3,#4,#5]#6{%
        \node[draw, drop shadow={opacity=.35}, minimum height=#1, minimum width=#2, %
        inner color=white, outer color=white, color=white!50!black] (#4) at #5 {}; %
        \node[anchor=#3,inner sep=2pt] at (#4.#3) {#6}; %
      }
      
      \def\tboxl[#1,#2,#3,#4,#5]#6{%
        \node[circle, draw, drop shadow={opacity=.35}, minimum height=#1, minimum width=#2, %
        inner color=orange!25, outer color=orange!48, color=orange!40!black] (#4) at #5 {}; %
        \node[anchor=#3,inner sep=2pt] at (#4.#3) {#6};%
      }
      
    \def\tboxr[#1,#2,#3,#4,#5]#6{%
      \node[circle, draw, drop shadow={opacity=.35}, minimum height=#1, minimum width=#2, %
      inner color=green!25, outer color=green!48, color=green!40!black] (#4) at #5 {}; %
      \node[anchor=#3,inner sep=2pt] at (#4.#3) {#6}; %
    }
    
        \def\tboxd[#1,#2,#3,#4,#5]#6{%
            \node[circle, draw, drop shadow={opacity=.35}, minimum height=#1, minimum width=#2, %
            inner color=red!25, outer color=red!48, color=red!40!black] (#4) at #5 {}; %
            \node[anchor=#3,inner sep=2pt] at (#4.#3) {#6};%
          }

      \def\entity[#1,#2]#3;{
        \node[draw,font=\tiny, drop shadow={opacity=.4,shadow xshift=0.04, shadow
          yshift=-0.04}, color=blue!30!black,fill=white,rounded corners=3] (#1) at #2 {#3};
      }
      
          \def\entityA[#1,#2]#3;{
              \node[draw,font=\tiny, drop shadow={opacity=.4,shadow xshift=0.04, shadow
                yshift=-0.04}, text width=2.6cm, color=blue!30!black,fill=white,rounded corners=3] (#1) at #2 {#3};
            }
      
         \def\entityy[#1,#2]#3;{
              \node[draw,font=\tiny, drop shadow={opacity=.4,shadow xshift=0.04, shadow
                yshift=-0.04}, color=blue!30!black,fill=white] (#1) at #2 {#3};
            }
      \def\isaedge[#1,#2,#3,#4];{ 
        \draw[-triangle 60,color=black!20!black,#4,fill=white] (#1) -- #3
        (#2);  
      }
        \def\isapath[#1,#2,#3,#4];{ 
              \draw[color=black!20!black,#4,fill=white] (#1) -- #3
              (#2);  
            }
            
        \def\isadashed[#1,#2,#3,#4];{ 
            \draw[arrow, -triangle 60, dotted,thick,color=black!40!black,#4,fill=white] (#1) -- #3
                          (#2);  
                 }
      
      \def\aboxl[#1,#2,#3,#4,#5]#6{%
        \node[draw, cylinder, alias=cyl, shape border rotate=90, aspect=#3, %
        minimum height=#1, minimum width=#2, outer sep=-0.5\pgflinewidth, %
        color=orange!40!black, left color=orange!70, right color=orange!80, middle
        color=white] (#4) at #5 {};%
        \node at #5 {#6};%
        \fill [orange!30] let \p1 = ($(cyl.before top)!0.5!(cyl.after top)$), \p2 =
        (cyl.top), \p3 = (cyl.before top), \n1={veclen(\x3-\x1,\y3-\y1)},
        \n2={veclen(\x2-\x1,\y2-\y1)} in (\p1) ellipse (\n1 and \n2); }
          
      \def\aboxr[#1,#2,#3,#4,#5]#6{%
        \node[draw, cylinder, alias=cyl, shape border rotate=90, aspect=#3, %
        minimum height=#1, minimum width=#2, outer sep=-0.5\pgflinewidth, %
        color=orange!50!black, left color=orange!50, right color=orange!60, middle
        color=white] (#4) at #5 {};%
        \node at #5 {#6};%
        \fill [orange!20] let \p1 = ($(cyl.before top)!0.5!(cyl.after top)$), \p2 =
        (cyl.top), \p3 = (cyl.before top), \n1={veclen(\x3-\x1,\y3-\y1)},
        \n2={veclen(\x2-\x1,\y2-\y1)} in (\p1) ellipse (\n1 and \n2); }
      
      \def\kbbox[#1,#2,#3,#4,#5]#6{
              \draw[dashed] node[draw,color=gray!50,minimum
              height=#1,minimum width=#2] (#4) at #5 {}; 
              \node[anchor=#3,inner sep=2pt] at (#4.#3)  {#6};
      }
      
      \def\soledge[#1,#2,#3,#4];{
              \draw[dashed,-latex,#4] (#1) -- #3 (#2);
      }

\usepackage{algorithmic}
\usepackage{algorithm}

\usepackage{epsfig}
\usepackage{latexsym}
\usepackage{newlfont}

\usepackage{fancyhdr} 
\usepackage[french]{todonotes} 

 % ...

\usepackage{moreverb}

\newcommand{\store}[5]{%
 \draw [fill=#1] (#2,#3) -- (#2,#5) to [out=270,in=155] (#2+#4/14-#2/14,#5-#4/14+#2/14) 
 to [out=335,in=180] (#2+#4/2-#2/2,#5-#5/5+#3/5) 
 to [out=0,in=205] (#4-#4/14+#2/14,#5-#4/14+#2/14) to [out=25,in=270] (#4,#5) -- (#4,#3) to [in=25,out=270] (#4-#4/14+#2/14,#3-#4/14+#2/14) to [in=0,out=205] (#2+#4/2-#2/2,#3-#5/5+#3/5) 
 to [in=335,out=180] (#2+#4/14-#2/14,#3-#4/14+#2/14) to [in=270,out=155] (#2,#3) -- (#2,#5) to [out=90,in=205] (#2+#4/14-#2/14,#5+#4/14-#2/14) to [out=25,in=180] (#2+#4/2-#2/2,#5+#5/5-#3/5) 
 to [out=0,in=155] (#4-#4/14+#2/14,#5+#4/14-#2/14) to [out=335,in=90] (#4,#5)
}
\newcommand{\symbolEnd}[4]{%
 \draw[#1,line width=3*#4] (#2+1.73*#4,#3+2*#4) -- (#2+1.73*#4,#3-2*#4) -- (#2-1.73*#4,#3) -- cycle;
 \filldraw [fill=#1,draw=#1] (#2+1.73*#4,#3+2*#4) circle [radius=#4];
 \filldraw [fill=#1,draw=#1] (#2+1.73*#4,#3-2*#4) circle [radius=#4];
 \filldraw [fill=#1,draw=#1] (#2-1.73*#4,#3) circle [radius=#4]
}

\newcommand{\feta}[0]{\textsf{FETA}\xspace}

\def\part{\mathcal{P}}

\makeatletter
\providecommand*{\toclevel@algorithm}{0}
\makeatother

\begin{document}
\title{Tracking Federated Queries in the Linked Data}
\author{Georges Nassopoulos, Patricia Serrano-Alvarado, \\Pascal Molli, Emmanuel Desmontils}


\institute{LINA Laboratory, Universit\'e de Nantes -- France\\ $firstName.lastName$@univ-nantes.fr}

\maketitle

\begin{abstract}

  Federated query engines allow data consumers to execute queries over
  the federation of Linked Data (LD). However, as federated queries
  are decomposed into potentially thousands of subqueries distributed
  among SPARQL endpoints, data providers do not know federated
  queries, they only know subqueries they process. %
  Consequently, unlike warehousing approaches, LD data providers have
  no access to secondary data. %
  In this paper, we propose \feta (\textsf{FE}derated query \textsf{T}r\textsf{A}cking), a query
  tracking algorithm that infers Basic Graph Patterns (BGPs) processed by a federation from a
  shared log maintained by data providers. %
  Concurrent execution of thousand subqueries generated by multiple
  federated query engines makes the query tracking process challenging
  and uncertain. %
  Experiments with Anapsid show that \feta is able to extract BGPs
  which, even in a worst case scenario, contain BGPs of original queries.
\end{abstract}



\input{Introduction.tex}

\input{problem.tex}
\input{algos.tex}
\input{experiments.tex}

\input{relatedWork.tex}

\input{conclusion.tex}

\bibliographystyle{abbrv}
\bibliography{myBib}  

\end{document}

%% file: Introduction.tex
\section{Introduction}

The federation of the Linked Data (LD) interlinks massive amounts of
data across the Web. Federated query
engines~\cite{fedex-fluid-schwarte-11,anapsid-vidal-11,splendid-federated-query-engine-11,hartig2009executing,quilitz2008querying}, allow data consumers to
query data residing in the federation in a transparent way as if they
were a single RDF graph.

Query engines split user's query into subqueries distributed among
SPARQL endpoints without revealing the whole federated query. Hence,
data providers do not know the complete federated query in which they
participate, they do not know which of their data are combined, when
and by whom. Consequently, data providers have a partial access to
secondary data~\cite{economic_seondary_data_2014,querying_aggregated_data_derived_data_GrumbachRT99},
unlike data warehousing approaches.

In this paper we propose \feta (\textsf{FE}derated query \textsf{T}r\textsf{A}cking), a query
tracking algorithm that computes original federated BGPs (Basic Graph Patterns) from shared
logs maintained by data providers. Concurrent execution of thousand
subqueries generated by multiple federated query engines makes the
query tracking process challenging and uncertain. To tackle this
problem, we developed a set of heuristics that links or unlinks
variables used in different subqueries of a join federated query. We experimented \feta over
concurrent execution of queries of the benchmark
FedBench~\cite{fedbench_Schmidt_dblp_2011}. Even in a worst case scenario,
\feta extracts BGPs that contain federated BGPs used in original queries.

The paper is organized as follows:
Section~\ref{sec:background_motivation} introduces a motivating
example and describes the scientific problem. Section~\ref{sec:feta}
presents \feta and heuristics for query deduction.
Section~\ref{sec:experiments} illustrates experimental results.
Section~\ref{sec:related_work} overviews some related work. Finally, conclusions and future work
are outlined in Section~\ref{sec:conclusion}.

%% file: problem.tex
\section{Background and Motivations}
\label{sec:background_motivation}

Given a SPARQL query and a federation defined as a set of SPARQL
endpoints, a federated query engine performs the following tasks~\cite{kossmann2000state,livreValduriez-2011}:
\begin{inparaenum}[(i)]
\item \emph{query decomposition}, normalizes, rewrites and simplifies
  queries;
\item \emph{data localization}, performs source selection among defined
  federation and rewrites the query into a distributed query;
\item \emph{global query optimization}, optimizes distributed query by
  rewriting an equivalent distributed query with various heuristics:
  minimizing intermediate results, minimizing number of calls to
  endpoints, etc. In Figure~\ref{fig:motivation_example}, we observe that two join operations will
  be executed in the federated query engine with data coming from
  subqueries sent to SPARQL endpoints;
\item \emph{distributed query execution},
  executes the optimized plan
  with physical operators available in federated query engines.
 \end{inparaenum}

\input{picture-motivation}

Federated query tracking infers federated queries from a shared
log maintained by data providers. We illustrate the general process in
Figure~\ref{fig:motivation_example}. A federated query engine executes
a federated query on a federation of SPARQL endpoints. \feta collects
the logs of a federation of data providers and infers BGPs used in
federated queries. By this way, data providers that collaborate have
access to secondary data. In our example, \feta allows NYTimes data provider to know which of his data are used in conjunction with DBPedia data.
 Query tracking can be applied to many federated query engines, in this paper we focus on tracking queries processed by the Anapsid~\cite{anapsid-vidal-11} federated query engine, with its join physical operators, nested loop  with filter options (\verb+nlfo+) and symmetric hash  (\verb+symhash+).

\begin{table}[ht!]
 \centering
 \scriptsize{
 \begin{tabular}{|l|l|l|}
 \hline 
 \textbf{Time} & \textbf{Subquery/Answer} & \textbf{Endpoint} \\
 \hline 

   11:24:19 & \textbf{(Subquery 1)} SELECT \verb+?pres+ WHERE \{ & DBPedia  \\
   & ?pres rdf:type dbpedia-owl:President  \} &  InstancesTypes \\
  
    \hline 
   11:24:23  & \textbf{(Answer)} \{ \{var: "pres", values: & DBPedia \\
   & "http://dbpedia.org/resource/Ernesto\_Samper,..., & InstancesTypes \\    
   & "http://dbpedia.org/resource/Shimon\_Peres,..., &  \\
   & http://dbpedia.org/resource/Barack\_Obama" \}  \} & \\
      \hline 
 \hline
  11:24:21  & \textbf{(Subquery 2)} SELECT ?party \verb+?pres+  WHERE \{  & DBPedia \\
  & ?pres dbpedia-owl:nationality dbpedia:United\_States . &  InfoBox \\
  &  ?pres dbpedia-owl:party ?party \} &  \\
  \hline 
  11:24:24 & \textbf{(Answer)} \{ \{var: "party", values:  & DBPedia  \\
  &  "http://dbpedia.org/resource/Democratic\_Party\_ & InfoBox  \\
  &  "\%28United\_States\%29,..., &   \\
  &  http://dbpedia.org/resource/Independent\_\%28politics\%29", & \\
  &  http://dbpedia.org/resource/Republican\_Party\_\%28US\%29" \}, & \\
  &  \{ var: "pres",  values:  &   \\
  & "\textbf{\emph{http://dbpedia.org/resource/Barack\_Obama}},..., &  \\  
  & \textbf{\emph{http://dbpedia.org/resource/Johnny\_Anders}},..., &  \\ 
  & \textbf{\emph{http://dbpedia.org/resource/Judith\_Flanagan\_Kennedy}},..." \}  \} & \\ 
  \hline
 \hline
  11:24:25  & \textbf{(Subquery 3)} SELECT \verb+?pres+ ?x ?page WHERE \{  & NYTimes \\
  & ?x nytimes:topicPage ?page .  &  \\
  & ?x owl:sameAs ?pres . FILTER  &  \\
  & ((\textbf{\emph{?pres=<http://dbpedia.org/resource/Barack\_Obama>}})  || &  \\
  & (\textbf{\emph{?pres=<http://dbpedia.org/resource/Johnny\_Anders>}}) || &  \\
  & (\textbf{\emph{?pres=<http://dbpedia.org/resource/Judith\_Flanagan\_Kennedy>}}),...) || &  \\
  & \}\} LIMIT 10000 OFFSET 0  &  \\
  \hline
  11:24:27 & \textbf{\textbf{(Answer)}} \{ \{var: "pres", values: & NYTimes \\
  & "http://dbpedia.org/resource/Barack\_Obama" \} & \\
  &  \{ var: "x", values: & \\
  &  "http://data.nytimes.com/47452218948077706853" \} & \\
  &  \{ var: "page", values: & \\
  &  "http://topics.nytimes.com/top/reference/timestopics/ & \\
  &  people/o/barack\_obama/index.html" \} & \\
 \hline
  
      \end{tabular}}
   \caption{Partial logs of DBPedia (InstancesTypes, InfoBox) and NYTimes.}
   \label{tab:query_logs}
\end{table}
 \input{picture-motivation2}

Figure~\ref{fig:motivation_example} illustrates how Anapsid processes query CD3 from FedBench\footnote{Prefixes for queries presented in this article are: \\ 
PREFIX dbpedia: <http://dbpedia.org/resource/>  \\
PREFIX dbpedia-owl: <http://dbpedia.org/ontology/> \\
PREFIX foaf: <http://xmlns.com/foaf/0.1/> \\
PREFIX geonames: <http://www.geonames.org/ontology\#> \\
PREFIX linkedMDB: <http://data.linkedmdb.org/resource/movie/> \\
PREFIX nytimes: <http://data.nytimes.com/elements/> \\
PREFIX owl: <http://www.w3.org/2002/07/owl\#> \\
PREFIX purl: <http://purl.org/dc/terms/> \\
PREFIX rdf: <http://www.w3.org/1999/02/22-rdf-syntax-ns\#> }~\cite{fedbench_Schmidt_dblp_2011}. The goal is to find all US presidents, their party membership and pages with news about them.    Table~\ref{tab:query_logs}, presents an extraction of federated log with traces of the CD3  execution. It contains some subqueries and associated answers\footnote{Query results are sent in \textit{JSON} format.} of endpoints. These traces correspond to subqueries sent by one query engine, identified by its IP address.

In Figure~\ref{fig:motivation_example} we see that Anapsid evaluates individually $tp1$ of CD3, at DBPedia-instances types. Subsequently, this query engine choses a \verb+nlfo+ implementation to join data retrieved from DBPedia-infobox and NYTimes. Consequently, Anapsid sends $tp2\Join tp3$ to DBPedia-infobox and stores intermediate results. Then, it calls NYTimes with several subqueries 
containing these intermediate results in filter options. This is confirmed in Table \ref{tab:query_logs}, where answers of subquery 2 from DBPedia-infobox are injected in the filter part of subquery 3 and sent to NYTimes. Anapsid iterates until all intermediate results are sent to NYTimes, in order to avoid reaching the endpoint's limit response. Finally, results of the \verb+nlfo+ implementation are joined locally at Anapsid with results of the first triple pattern's evaluation, using the \verb+symhash+ operator.  

Operator \verb+nlfo+ is represented by an arrow because the order of the join is deduced from logs, i.e., it is possible to know in which direction the nested loop is made. But, \verb+symhash+ is represented by a dash line because it is impossible to know the order of the join made locally by the query engine.

It is clear that a single federated query can generate many subqueries sent to endpoints according to physical join operators. However, such behavior can be tracked if endpoints collaborate and BGPs from the federated query can be inferred. 

Federated query tracking is challenging if many federated queries having
common join conditions are executed concurrently. The most challenging case
is when, in addition, queries are sent by the same query engine. We
propose some heuristics to separate subqueries belonging to different
federated queries sent by the same query
engine. Figure~\ref{fig:motivation_example2} shows the concurrent
execution of queries CD3 and CD4 sent by query engine $QE_{i}$. These
queries have common variable \verb+?x+. When logs are federated, this
variable joins BGPs of both queries. 
 
\textbf{Problem statement.} Given a federated log containing
independent subqueries, link subqueries on their common join conditions, i.e., variable, IRI or literal, if they participate to the same federated query and deduce BGPs processed by
the federation. The desired output is a set of BGPs indicating
\begin{inparaenum}[(i)]
 \item  which endpoints evaluated which triple
   patterns,
 \item whom issued the federated query, and
 \item in which period of time the deduced BGP was processed.
 \end{inparaenum}
 \newline
At the bottom of Figures \ref{fig:motivation_example} and \ref{fig:motivation_example2} there are deduced BGPs corresponding to queries appearing at the top of these Figures. Information about which endpoints collaborate with which triple pattern, the federated query engine that issued the query, and the time period where the queries were processed appear too.


%% file: picture-motivation.tex
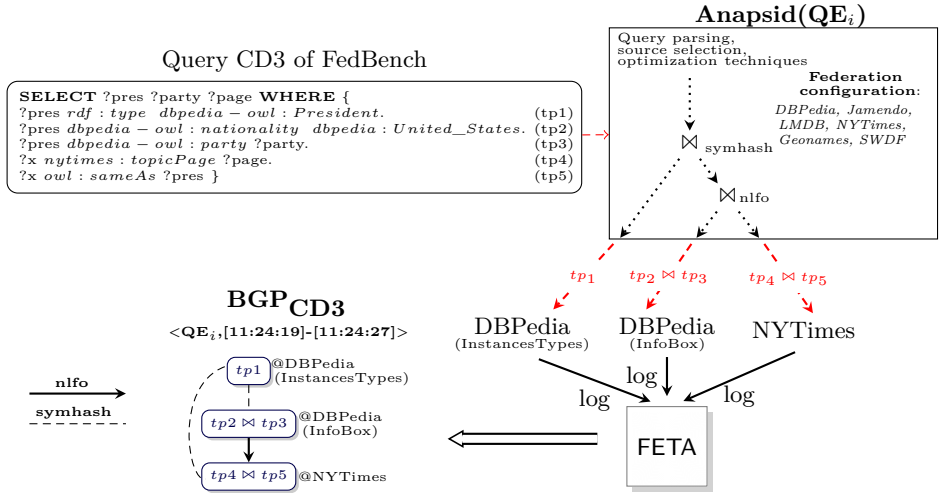
\begin{figure}[ht!]
\centering
\begin{tikzpicture}
\centering
\tikzstyle{vecArrow} = [thick, decoration={markings,mark=at position
   1 with {\arrow[semithick]{open triangle 60}}},
   double distance=3pt, shorten >= 5.5pt,
   preaction = {decorate},
   postaction = {draw,line width=2pt, white,shorten >= 4.5pt}]

\node(client)at(-0.4,0)[draw, align=left, rounded corners, text width=7.34cm, 
font=\tiny,label=above:Query CD3 of FedBench]{
\begin{tabular}{lc}
\textbf{SELECT} ?pres ?party ?page \textbf{WHERE} \{ & \\
?pres $rdf:type~$ $dbpedia-owl:President$.  & (tp1) \\
?pres $dbpedia-owl:nationality~$ $dbpedia:United\_States$. & (tp2) \\
?pres $dbpedia-owl:party$ ?party. &(tp3) \\
?x $nytimes:topicPage$ ?page. & (tp4) \\
?x $owl:sameAs$ ?pres  \} & (tp5)
  \end{tabular}}; 
   
        \entity[A3,(-1,-3.8)] {$tp2 \Join tp3$};
        \entity[B3,(-1,-4.5 )] {$tp4 \Join tp5$};
        \entity[C3,(-1,-3.1)] {$tp1$};
        \path[arrow,thick] (A3) edge (B3);
        \path[densely dashed] (C3) edge (A3);
                     
        \node(94)[] at(-1.2,-3){};
        \node(95)[] at(-1.45,-4.5){};
        \draw[densely dashed] (94) .. node[above] {} controls (-2,-3.3) and (-1.8,-4.6) ..  (95); 
            
\tboxu[30,30,north east,t4,(4.5,-4.1)] {};
  
\node(anapsid)[draw,text
width=4.1cm,minimum height=2.8cm] at(5.9,0.05){};

\node(0)[] at(4.8,0.9){};
\node(1)[] at(4.8,-0.1){$\Join$};

\node(2)[] at(5.3,-0.8){$\Join$};

\node(3)[] at(3.8,-1.45){};

\node(5)[] at(4.8,-1.45){};

\node(4)[] at(5.8,-1.45){};

\node(6)[text width=12em,minimum height=2em,font=\tiny] at(5.8,1.3){Query parsing,}; 
\node(7)[text width=12em,minimum height=2em,font=\tiny] at(5.8,1.15){source selection,};
\node(8)[text width=12em,minimum height=2em,font=\tiny] at(5.8,1){optimization techniques};
\node(27)[text width=15em,minimum height=2em,font=\tiny] at(8.8,0.8){\textbf{Federation}};
\node(45)[text width=15em,minimum height=2em,font=\tiny] at(8.6,0.6){\textbf{ configuration}:};
\node(28)[text width=15em,minimum height=2em,font=\tiny] at(8.35,0.35){\textit{DBPedia, Jamendo,}};
\node(29)[text width=15em,minimum height=2em,font=\tiny] at(8.4,0.15){\textit{LMDB, NYTimes,}};
\node(44)[text width=15em,minimum height=2em,font=\tiny] at(8.4,-0.05){\textit{Geonames, SWDF}};

\node(41)[] at(-1.2,-1.9){};
\node(42)[] at(-0.3,-1.9){};

\node(11)[font=\tiny] at(5.45,-0.15){symhash};
\node(12)[font=\tiny] at(5.65,-0.8){nlfo};
\node(13)[] at(1.05,-4){};

\node(14)[] at(4.9,-1.3){};
\node(15)[] at(4.15,-2.4){};
\node(16)[] at(3.9,-1.3){};
\node(17)[] at(2.9,-2.4){};
\node(18)[] at(5.7,-1.3){};
\node(19)[] at(6.5,-2.45){};
\node(20)[] at(4.5,-2.5){DBPedia};
\node(21)[text width=12em,minimum height=2em,font=\tiny] at(3.65,-2.75){(InstancesTypes)};
\node(22)[] at(2.6,-2.5){DBPedia};
\node(27)[] at(6.3,-2.55){NYTimes};
\node(23)[text width=12em,minimum height=2em,font=\tiny] at(5.95,-2.75){(InfoBox)};
\node(24)[text width=12em,minimum height=2em,font=\tiny] at(1.24,-3){@DBPedia};
\node(54)[text width=12em,minimum height=2em,font=\tiny] at(1.28,-3.2){(InstancesTypes)};
\node(25)[text width=12em,minimum height=2em,font=\tiny] at(1.6,-4.5){@NYTimes};
\node(26)[text width=12em,minimum height=2em,font=\tiny] at(1.6,-3.7){@DBPedia};
\node(56)[text width=12em,minimum height=2em,font=\tiny] at(1.65,-3.9){(InfoBox)};
\node(39)[] at(6,1.6){\textbf{Anapsid(QE$_{i}$)}};
\node(30)[] at((2.7,-2.9){};
\node(31)[] at(4.5,-2.8){};
\node(32)[] at(6.3,-2.8){};
\node(33)[] at(4.5,-4.1){\feta};
\node(34)[] at(4.4,-3.55){};
\node(35)[] at(4.5,-3.55){};
\node(36)[] at(4.6,-3.55){};
\node(37)[] at(-0.5,-2.2){\textbf{BGP$_{\textbf{CD3}}$}};
\node(38)[text width=20em,minimum height=2em,font=\tiny] at(1.2,-2.6){\textbf{<QE$_{i}$,[11:24:19]-[11:24:27]>}};
\node(39)[] at(3.7,-4){};
\node(40)[] at(1.5,-4){};
\path[arrow,thick] (30) edge node[below] {log} (34); 
\path[arrow,thick] (31) edge node[left] {log} (35);
\path[arrow,thick] (32) edge node[below] {log} (36);

  
\path[arrow,dotted] (0) edge (1)
 					(1) edge (3)
                        edge (2)
                    (2) edge (4)
                        edge (5);

  \path[->]
    (client) edge[red,densely dashed] (anapsid);
  
  \path[arrow](14) edge[red,densely dashed] node[fill=white,font=\tiny]{$tp_2 \Join tp_3$} (15);
  \path[arrow](16) edge[red,densely dashed] node[fill=white,font=\tiny]{$tp_1$} (17);
  \path[arrow](18) edge[red,densely dashed] node[fill=white,font=\tiny]{$tp_4 \Join tp_5$} (19);
 
 \draw[vecArrow] (39) to (40);
 
 
    \node(80)[] at(-4,-3.4){};
     \node(81)[] at(-2.5,-3.4){};
     \node(82)[] at(-4,-3.8){};
     \node(83)[] at(-2.5,-3.8){};
     
       \node(84)[text width=12em,minimum height=2em,font=\tiny] at(-1.6,-3.25){\textbf{nlfo}};
         \node(85)[text width=12em,minimum height=2em,font=\tiny] at(-1.85,-3.65){\textbf{symhash}};
     \path[arrow,thick] (80) edge (81);
     		              
     \path[densely dashed] (82) edge (83);
     
\end{tikzpicture}
 \caption{Query processing and \feta's deduction for CD3.}
	  \label{fig:motivation_example}
	\end{figure}

%% file: picture-motivation2.tex
\begin{figure}[ht!]
\begin{tikzpicture}
\tikzstyle{vecArrow} = [thick, decoration={markings,mark=at position
   1 with {\arrow[semithick]{open triangle 60}}},
   double distance=3pt, shorten >= 5.5pt,
   preaction = {decorate},
   postaction = {draw,line width=2pt, white,shorten >= 4.5pt}]

\node(client)at(-0.73,0)[draw, align=left, rounded corners, text width=7.34cm, 
font=\tiny,label=above:Query CD3 of FedBench]{
\begin{tabular}{lc}
\textbf{SELECT} ?pres ?party ?page \textbf{WHERE} \{  & \\
?pres $rdf:type~$ $ dbpedia-owl:President$  .  & (tp1) \\
?pres $dbpedia-owl:nationality~$ $dbpedia:United\_States$ . & (tp2) \\
?pres $dbpedia-owl:party$ ?party . &(tp3) \\
?x $nytimes:topicPage$ ?page . & (tp4) \\
?x $owl:sameAs$ ?pres  \} & (tp5)
  \end{tabular}}; 
  
\node(queryEngine)[draw,text width=7.1cm,minimum height=1.8cm] at(1.68,-2.3){};
    
\node(client2)at(5.42,0)[draw, align=left, rounded corners, text width=4.35cm, 
font=\tiny,label=above:Query CD4 of FedBench]{
\begin{tabular}{lc}
\textbf{SELECT} ?actor ?news \textbf{WHERE} \{ & \\
?film purl:title 'Tarzan' .  & ($tp_{1}$)  \\
?film linkedMDB:actor ?actor .  & ($tp_{2}$) \\
?actor owl:sameAs ?x  .  & ($tp_{3}$)  \\
?y owl:sameAs ?x .  & ($tp_{4}$)  \\
?y nytimes:topicPage ?news \}   &  ($tp_{5}$) 
  \end{tabular}};
  
   \entity[A3,(-1.4,-7.8)] {$tp2 \Join tp3$};
   \entity[B3,(-1.4,-8.8)] {$tp4 \Join tp5$};
   \entity[C3,(-1.4,-6.8)] {$tp1$};

         
         \node(94)[] at(-1.6,-6.7){};
         \node(95)[] at(-1.9,-8.8){};
         
    \draw[densely dashed] (94) .. node[above] {} controls (-2.6,-7.8) and (-2.3,-9) ..  (95); 
             
   \entity[A4,(4.25,-6.8)] {$tp_1 \Join tp_2 \Join tp_3$};
   \entity[B4,(4.25,-7.8)] {$tp_5 \Join tp_4$};
\tboxu[30,30,north east,t4,(1.95,-5.95)] {};

 \path[arrow,thick]  (A3) edge (B3)
  					 (A4) edge (B4);
  					 
\path[densely dashed,thick](A3) edge (C3);

width=3cm,minimum height=2cm] at(1.1,-2.5){};
\node(0)[] at(1.05,-1.4){};
\node(60)[] at(1.15,-1.45){};
\node(1)[] at(1.05,-2.1){$\Join$};

\node(2)[] at(1.8,-2.6){$\Join$};

\node(3)[] at(0,-3.3){};

\node(5)[] at(2.3,-3.3){};

\node(4)[] at(1.4,-3.3){};

width=2.6cm,minimum height=2cm] at(4.5,-2.5){};
\node(50)[] at(5.5,-1.5){};
\node(51)[] at(4.6,-2.5){$\Join$};
\node(52)[font=\tiny] at(4.95,-2.5){nlfo};

\node(53)[] at(4.1,-3.3){};
\node(63)[] at(4.22,-3.1){};
\node(56)[] at(3.4,-4.35){};

\node(54)[] at(6.5,-4.15){};
\node(55)[] at(4.6,-1.4){};
\node(65)[] at(4.5,-1.5){};
\node(57)[] at(5.15,-3.3){};
\node(67)[] at(5.05,-3.15){};
\node(58)[] at(5.7,-4.3){};
\node(6)[text width=12em,minimum height=2em,font=\tiny] at(0.05,-1.6){Query parsing,}; 
\node(7)[text width=12em,minimum height=2em,font=\tiny] at(0.05,-1.8){source selection,};
\node(8)[text width=12em,minimum height=2em,font=\tiny] at(0.05,-2.0){optimization techniques};
\node(9)[] at(0.2,-1.7){};
\node(10)[] at(1,-1.7){};

\node(27)[text width=15em,minimum height=2em,font=\tiny] at(0.95,-2.3){\textbf{Federation}};
\node(87)[text width=15em,minimum height=2em,font=\tiny] at(0.75,-2.5){\textbf{configuration}:};
\node(28)[text width=15em,minimum height=2em,font=\tiny] at(0.57,-2.7){\textit{DBPedia, Jamendo,}};
\node(29)[text width=15em,minimum height=2em,font=\tiny] at(0.55,-2.9){\textit{LMDB, NYTimes,}};
\node(89)[text width=15em,minimum height=2em,font=\tiny] at(0.55,-3.1){\textit{Geonames, SWDF}};

\node(11)[font=\tiny] at(1.7,-2.1){symhash};
\node(12)[font=\tiny] at(2.2,-2.6){nlfo};
\node(13)[] at(1.05,-4){};

\node(14)[] at(2.15,-3.1){};
\node(15)[] at(2.85,-4.3){};
\node(16)[] at(0.15,-3.1){};
\node(17)[] at(-0.9,-4.3){};
\node(18)[] at(1.5,-3.1){};
\node(19)[] at(0.85,-4.3){};
\node(20)[] at(-1.15,-4.37){DBPedia};
\node(21)[text width=12em,minimum height=2em,font=\tiny] at(-0.10,-4.62){(InstancesTypes)};

\node(22)[] at(1.15,-4.37){DBPedia};
\node(23)[text width=12em,minimum height=2em,font=\tiny] at(2.6,-4.62){(InfoBox)};

\node(24)[text width=12em,minimum height=2em,font=\tiny] at(0.9,-6.7){@DBPedia};
\node(84)[text width=12em,minimum height=2em,font=\tiny] at(0.85,-6.9){(InstancesTypes)};
\node(25)[text width=12em,minimum height=2em,font=\tiny] at(1.2,-8.8){@NYTimes};

\node(26)[text width=12em,minimum height=2em,font=\tiny] at(1.2,-7.7){@DBPedia};
\node(86)[text width=12em,minimum height=2em,font=\tiny] at(1.25,-7.9){(InfoBox)};
\node(74)[text width=12em,minimum height=2em,font=\tiny] at(7.2,-6.8){@LMDB};
\node(76)[text width=12em,minimum height=2em,font=\tiny] at(6.85,-7.8){@NYTimes};
\node(39)[] at(-3.2,-2.2){\textbf{Anapsid(QE$_{i}$)}};
\node(27)[] at(3.2,-4.42){NYTimes};
\node(28)[] at(5.4,-4.42){LMDB};

\node(30)[] at(-1.3,-4.8){};
\node(31)[] at(1.1,-4.65){};
\node(32)[] at(3.0,-4.6){};
\node(33)[] at(1.95,-5.95){\feta};
\node(34)[] at(1.6,-5.4){};
\node(35)[] at(1.95,-5.5){};
\node(36)[] at(1.95,-5.48){};
\node(42)[] at(5.6,-4.65){};
\node(46)[] at(2.27,-5.4){};

\node(37)[] at(-0.6,-5.9){\textbf{BGP$_{\textbf{CD3}}$}};
\node(38)[text width=20em,minimum height=2em,font=\tiny] at((1,-6.3){\textbf{<QE$_{i}$,[11:24:19]-[11:24:27]>}};
\node(39)[] at(1.4,-5.9){};
\node(40)[] at(0.8,-5.9){};
\node(69)[] at(2.55,-5.9){};
\node(70)[] at(3.15,-5.9){};
\node(71)[] at(4.7,-5.9){\textbf{BGP$_{\textbf{CD4}}$}};
\node(72)[text width=20em,minimum height=2em,font=\tiny] at(6.3,-6.3){\textbf{<QE$_{i}$,[11:24:23]-[11:24:28]>}};

 \path[arrow,thick] (30) edge  node[below,font=\tiny]{log} (34)
  				     (31) edge node[left,font=\tiny]{log} (35)
  					 (32) edge node[right,font=\tiny]{log} (36)
  					 (42) edge node[below,font=\tiny]{log} (46);
  					 
\path[arrow,dotted] (0) edge (1)
 					(1) edge (3)
                        edge (2)
                    (2) edge (4)
                        edge (5)
                    (55) edge (51)
                    (51) edge (53)
                         edge (57);

  \path[->]
    (client) edge[red,densely dashed] node[fill=white,left]{query}(60);
  
  \path[->]
      (client2) edge[red,dashed] node[fill=white,right]{query}(65);
  
  \path[arrow](14) edge[red,dashed] node[fill=white,font=\tiny]{$tp_2 \Join tp_3$} (15);
  \path[arrow](16) edge[red,dashed] node[fill=white,font=\tiny]{$tp_1$} (17);
  \path[arrow](18) edge[red,dashed] node[fill=white,font=\tiny]{$tp_4 \Join tp_5$} (19);
 
   \path[arrow](67) edge[red,dashed] node[fill=white,font=\tiny]{$tp_1 \Join tp_2 \Join tp_3$} (58);
   \path[arrow](63) edge[red,dashed] node[fill=white,font=\tiny]{$tp_5 \Join tp_4$} (56);
   
 \draw[vecArrow] (39) to (40);
  \draw[vecArrow] (69) to (70);
 
  \draw[vecArrow] (9) to (10);
   \node(80)[] at(1.3,-6.9){};
    \node(81)[] at(2.6,-6.9){};
    \node(82)[] at(1.3,-7.3){};
    \node(83)[] at(2.6,-7.3){};
    
      \node(84)[text width=12em,minimum height=2em,font=\tiny] at(3.6,-6.75){\textbf{nlfo}};
        \node(85)[text width=12em,minimum height=2em,font=\tiny] at(3.4,-7.15){\textbf{symhash}};
    \path[arrow,thick] (80) edge (81);
    		              
    \path[densely dashed] (82) edge (83);
   
\end{tikzpicture}
 \caption{Query processing and \feta's deduction for CD3, CD4 concurrent execution.}
	  \label{fig:motivation_example2}
	\end{figure}
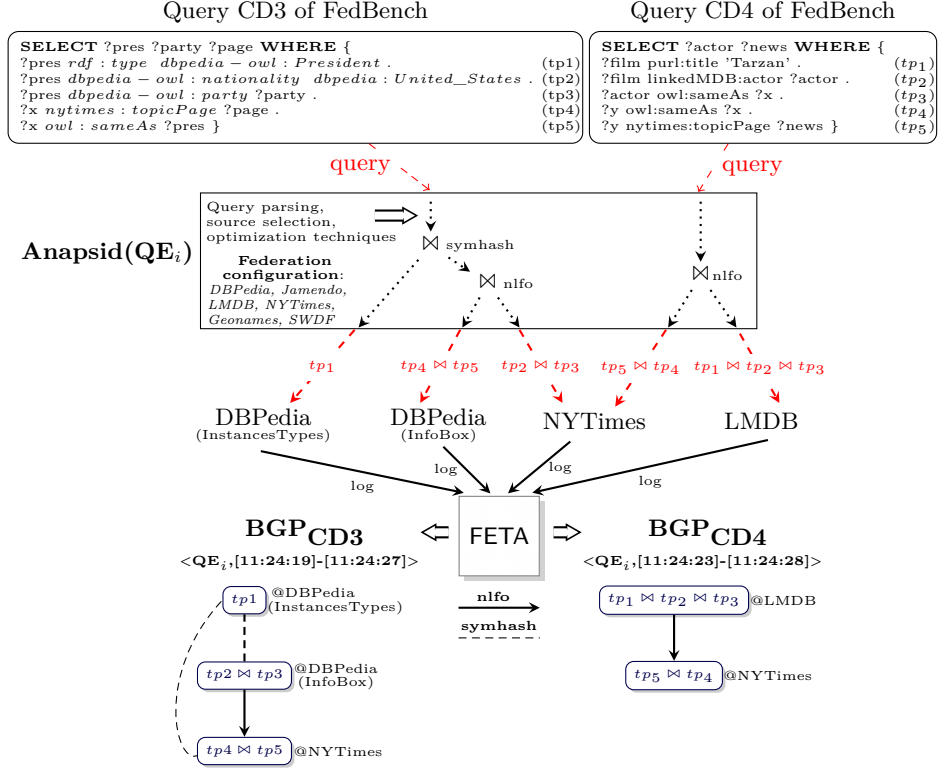

%% file: algos.tex
\section{\feta: FEderated query TrAcking}
\label{sec:feta}

\begin{figure}[ht!]
 	      \centering
  \begin{tikzpicture}[
  node distance=16mm,
      start chain=going right,
      diagram item/.style={
          on chain,
          join
      }
  ]
       
 \tikzstyle{vecArrow} = [thick, decoration={markings,mark=at position
    1 with {\arrow[semithick]{open triangle 60}}},
    double distance=3pt, shorten >= 5.5pt,
    preaction = {decorate},
    postaction = {draw,line width=2pt, white,shorten >= 4.5pt}]
 
  \node[database] (3) at (-0.5,0) {};
 \node (4) [rectangle, minimum width=0.2cm, minimum height=0.5cm, text centered, draw=black]at(0.9,0) {T$_\Join$};
  \node (5) [rectangle, minimum width=1.5cm, minimum height=0.4cm, text centered, draw=black] at (2.4,0){};
 \node (6) [rectangle, minimum width=1.6cm, minimum height=0.6cm, text centered, draw=black] at(4.35,0) {};
 \node (7) [rectangle, minimum width=0.8cm, minimum height=0.4cm, text centered, draw=black] at(6,0) {};
 \node (8) [rectangle, minimum width=1.7cm, minimum height=0.6cm, text centered, draw=black] at(7.75,0) {};
 \node (9) [rectangle, minimum width=1.5cm, minimum height=0.4cm, text centered, draw=black] at(9.82,0) {};
 \node(10)[] at(-0.48,-1.6){};
 \node(11)[] at(9.8,-1.3){};
 \node(12)[] at(5.1,-1.2){};
 \node(13)[] at(7,-1.2){};
 \node(14)[] at(5.1,-1.55){};
 \node(15)[] at(7,-1.55){};

 \draw[vecArrow] (10) to (3);
 \draw[vecArrow] (9) to (11);
 \node[] at(-0.5,0.15){Federated}; 
 \node[] at(-0.5,-0.2){Log};
 \node[text width=7em,minimum height=2em,font=\tiny] at(2.85,-0){Same queries}; 
 \node[text width=7em,minimum height=2em,font=\tiny] at(5.05,0.15){Common}; 
 \node[text width=7em,minimum height=2em,font=\tiny] at(4.75,-0.1){join condition};
 \node[text width=7em,minimum height=2em,font=\tiny] at(6.8,0){NLFO}; 
 \node[text width=7em,minimum height=2em,font=\tiny] at(8.1,0.1){Same Concept/}; 
 \node[text width=7em,minimum height=2em,font=\tiny] at(8.45,-0.15){Same As}; 
 \node[text width=7em,minimum height=2em,font=\tiny] at(10.25,0){Not Null Join};
 \node[] at(-0.5,-1.7){\feta input}; 
 \node[] at(9.7,-1.4){\feta output}; 
  \node[font=\tiny] at(-0.5,-1.95){(subqueries/answers)}; 
  \node[font=\tiny] at(9.7,-1.65){(BGPs)}; 
 \node[] at(6,-1){subqueries};
 \node[] at(6,-1.4){answers};    
 
 \path[arrow,thin] (3) edge (4)
 		          (4) edge (5)
 		          (5) edge (6)
 		          (6) edge (7)
 		          (7) edge (8) 
 		          (8)  edge (9)
 		          (12) edge (13);
 		              
 \path[arrow,dotted] (14) edge (15);
                     
 \draw[-latex, dotted, thick] (3) .. node[above] {} controls (1.1,1) and (9.6,1) ..  (9); 
 \draw[-latex, dotted, thick] (3) .. node[above] {} controls (1.2,0.9) and (7,0.9) ..  (8); 
  \draw[-latex, dotted, thick] (3) .. node[above] {} controls (1.3,0.8) and (5.4,0.84) ..  (7);  
   \end{tikzpicture}
  \caption{Workflow processing of \feta's heuristics.}
 	  \label{fig:feta_architecture}
\end{figure}
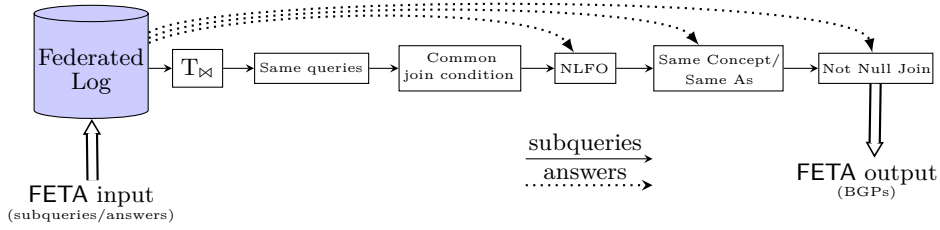

Figure~\ref{fig:feta_architecture} describes how \feta processes a
federated log. A federated log is a sequence of subqueries with
answers as described in Table~\ref{tab:query_logs}. The goal is to
link different subqueries participating in the same join, in order to
reconstruct federated BGPs. \emph{T$_\Join$}, \emph{Same queries} and \emph{Common join condition}, allow to join subqueries'
BGPs. \emph{NLFO}, \emph{Same Concept/Same As} and \emph{Not Null Join}
verify joins and potentially unlink BGPs.

\emph{T$_\Join$} identifies subqueries of the same time interval
that will be analyzed together. For instance, all subqueries in Table
\ref{tab:query_logs}, are captured in the interval of few seconds. The
challenge is to choose the appropriate time interval. A small window
may separate subqueries pertaining to the same federated query. A
large window may join subqueries of different federated queries.

\emph{Same queries} merges identical subqueries but also
subqueries differing only in their offset values. Same queries are
sent twice to the same endpoint to be sure obtaining an answer and to
different endpoints in order to have complete answers. For instance,
every query in Table \ref{tab:query_logs} is sent twice consecutively
to each selected endpoint. Additionally in Figure~\ref{fig:feta_cd_output3}, we observe that the first two triple patterns of CD6 are evaluated separately to different endpoints, with the aim of having complete answers. Similar queries with different offsets, on the other hand, are sent to avoid reaching the endpoint's limit response. For instance, the subqueries below which are sent to Geonames for evaluating query CD7, are merged by \feta and considered as a unique subquery without the part limit/offset:

\begin{tikzpicture}
\node [draw, text width=9.5cm] at(0,1.7)  {SELECT ?location ?news WHERE \{ \\  ?y <http://data.nytimes.com/elements/topicPage> ?news \\ ?y <http://www.w3.org/2002/07/owl\#sameAs> ?location \} \\  LIMIT 10000 OFFSET 0 \\};

\node [draw, text width=9.5cm] at (0,-0.2){SELECT ?location ?news WHERE \{ \\  ?y <http://data.nytimes.com/elements/topicPage> ?news \\ ?y <http://www.w3.org/2002/07/owl\#sameAs> ?location \} \\  LIMIT 10000 OFFSET 10000 };
\end{tikzpicture}

\emph{Common join condition}, joins BGPs of queries having common
projected variables or triple patterns with common IRI/literal on their subjects or objects. We are aware that, in general, subqueries are joined on their common projected variables. However, we consider also IRIs and literals, even if it can produce some noise on our deduction approach, because they are used in some cases as a common join condition. For instance, in Figure~\ref{fig:feta_cd_output1}, the IRI \emph{dbpedia:Barack\_Obama}, is a join condition between triple patterns of CD2. We presume that subqueries with common join condition, closely in time, may be joined locally at the query engine using the symmetric hash operator. For instance, in Table
\ref{tab:query_logs}, all subqueries have variable \verb+?pres+ in common and thus we
suppose they are joined at Anapsid. 

\emph{Nested Loop with Filter Options (NLFO)}, verifies if BGPs, joined in
the previous heuristic, were executed with a nested loop operator. In
particular, we group queries varying only in their filter
values, if these values are contained in answers of a previously evaluated subquery, with which we confirm that they are joined. %
For instance, in Table \ref{tab:query_logs} we identify that
filter values of subquery 3 correspond to answers of variable \verb+?pres+, e.g., \emph{<http://dbpedia.org/resource/Barack\_Obama>}, for subquery
2. This certifies a \verb+nlfo+ between subqueries 2 and 3, discarding a
global \verb+symhash+ join among the three subqueries.

\emph{Same Concept/Same As}, verifies if answers of joined queries
correspond to same concepts or concepts related with a \textit{sameAs}
property.\footnote{Note that we do not consider generic concepts for this heuristic, e.g., \newline \textit{<http://www.w3.org/2002/07/owl\#Thing>}.} If this is not the case, concerned BGPs are unlinked. %
For instance, in Table \ref{tab:query_logs}, answers of the triple
pattern of subquery 1 have the same concept with the second triple
pattern of subquery 3, for variable \verb+?pres+, i.e.,
\emph{<http://dbpedia.org/ontology/President>}.

\emph{Not Null Join}, verifies if a join returns an empty set of
answers. If this is the case, concerned BGPs are unlinked. For
instance, in Table \ref{tab:query_logs}, triple patterns in subqueries
1 and 3 have a common value for projected variable \verb+?pres+, i.e.,
\emph{<http://dbpedia.org/resource/Barack\_Obama>} and therefore they remain linked in the same BGP.

%% file: experiments.tex
\section{Experiments}
\label{sec:experiments}

We analyzed the collection of 7 federated queries of Cross Domain (CD)
of the benchmark FedBench~\cite{fedbench_Schmidt_dblp_2011}. Datasets
are those concerned by these queries: DBPedia, Jamendo, LMDB, NYTimes,
SWDF and Geonames. Virtuoso
OpenLink\footnote{http://virtuoso.openlinksw.com/} 6.1.7 is hosting
SPARQL endpoints. We used Anapsid 2.7 as federated query engine with
the cache disabled. Answers of endpoints to the subqueries they
received, are captured with tcpdump
4.5.1\footnote{http://www.tcpdump.org/}. \feta is implemented in Java
1.7 and the collected federated log is stored in a CouchDB
database\footnote{http://couchdb.apache.org/}.

We evaluated \feta under two configurations. In the first
configuration, one Anapsid client processes all federated queries
sequentially with a delay between each query. In the second one, one
Anapsid client processes all queries concurrently, each one into an
individual thread. Executing queries concurrently from a single client
is clearly a worst case scenario for \feta because the IP address of the client
cannot be used to split subqueries of the federated log. For the scope of
this paper, we suppose that all endpoints concerned by federated
queries share their logs.

With the first configuration, \feta reconstructs correctly all
federated BGPs of the CD collection. \footnote{Each deduced BGP corresponds to the federated query, once simplified and rewritten by the query engine at the \emph{query decomposition} phase.} We focus now on the
second experiment. In this case, Anapsid produces 529
subqueries.\footnote{Note that we subsequently remove ASKs and consider only SELECT subqueries.} Size of queries and answers logs are of 300KB and 14MB, respectively.

\begin{table}[ht]
  \footnotesize
  \centering
  \begin{tabular}{|l|l|}
     \hline
      \bf{\feta Heuristic}  & \bf{Number of produced BGPs} \\
     \hline
     Same queries & 109 \\
     \hline
     Common join condition & 1 \\
     \hline
      NLFO &  1 \\
     \hline
     Same Concept/Same As & 2 \\
     \hline
     Not Null Join & 4 \\
     \hline  
  \end{tabular}
  \caption{Number of BGP's produced by heuristic.}
  \label{tab:evaluation}
\end{table}

Table \ref{tab:evaluation} shows the number of BGPs produced after each heuristic following the \feta execution workflow shown in Figure~\ref{fig:feta_architecture}. \feta processes this federated log in approximately 90 seconds. $T_\Join$ is not significant here, we consider it big enough to cover the execution of all federated queries of CD. Initial log contains 238 SELECT subqueries as they are unlinked, there are 238 BGPs. \emph{Same queries} heuristic removes or merges more than 60\% of subqueries and their respective answers, producing 109 BGPs. \emph{Common join condition} produces a single BGP because chaining among queries. \emph{NLFO} confirms joins, by identifying the injection of answers from a subquery into subqueries which vary only in their filter values. \emph{Same Concept/Same As} heuristic unlinks some joins and certifies others. \emph{Not Null Join} certifies that a symmetric hash is certainly possible, because an intersection of answers on a common projected variable of every two subqueries.


\input{picture_feta_deduction.tex}

Figures~\ref{fig:feta_cd_output1}-\ref{fig:feta_cd_output4} present BGPs extracted by \feta for all concurrently executed CD queries. Note that query plans established by Anapsid for each query, may differ depending on endpoints availability and when operators are blocked. Ideally, \feta should reconstruct 8 BGPs. The CD collection consists of 7 queries but CD1 is a union query normally decomposed in 2 BGPs. \feta extracted 4 BGPs containing the 8 original BGPs. Even if this result is not precise, extracted BGPs with endpoints' information give valuable information to data providers bout how their data are processed and (potentially) joined with other endpoints. In the following paragraphs, we explain how \feta deduces each particular BGP.
 
Figure~\ref{fig:feta_cd_output1} describes how \feta processes federated queries CD1, CD2 and CD3. CD1 is composed of two BGPs separated by a union, which we expect to identify individually. In fact, these two BGPs were deduced as a single BGP because they have a common IRI, \emph{dbpedia:Barack\_Obama} and also share common answers for both variables \verb+?predicate+ and \verb+?object+ which concern \textit{Barack Obama}. Next, we observe that CD1, CD2, and CD3 were grouped in one BGP. CD1 and CD2 were not separated because of the common IRI  \emph{dbpedia:Barack\_Obama}, which is actually also a join condition between triple patterns of CD2. BGPs of CD2 and CD3 were not separated because results of CD2 are included in CD3 for their common triple pattern, $?x \ nytimes:topicPage \ ?page$, but also for the other triple patterns of CD2.
 

In Figures~\ref{fig:feta_cd_output2} and \ref{fig:feta_cd_output3}, we can see that BGPs of CD5 and CD6 were well reconstructed. CD4 and CD5 are linked in the same BGP with \emph{Same Concept/Same As}, as both concern films. Subsequently, \emph{Not Null Join} separated CD4 from CD5 on the content of the \verb+?film+ variable, as CD4 concerns films related to Tarzan while CD4 concerns films of Italian directors. In a similar way, CD6 and CD7 are linked with \emph{Same Concept/Same As}, as both concern localizations but they were separated because they have no common answer for variable \verb+?location+ as the first concerns the Federal Republic of Germany and the second California, USA.

Figure~\ref{fig:feta_cd_output4} shows \feta's deduced BGP, grouping CD4 and CD7. On the other hand CD4 and CD7 share common concepts but also answers because, for the currently employed heuristics, we infer that these two queries share the same triple pattern $?y \ nytimes:topicPage \ ?news$. 

From this experiments we conclude that (i) it is possible to reconstruct precise federated BGPs if federated queries are different enough, and (ii) reconstructed BGPs contain all original BGPs, i.e., false joins are not deduced.

%% file: picture_feta_deduction.tex
\begin{figure}[ht!]
\begin{tikzpicture}
\tikzstyle{vecArrow} = [thick, decoration={markings,mark=at position
   1 with {\arrow[semithick]{open triangle 60}}},
   double distance=3pt, shorten >= 5.5pt,
   preaction = {decorate},
   postaction = {draw,line width=2pt, white,shorten >= 4.5pt}]
\def\kbbox[#1,#2,#3,#4,#5]#6{
        \draw[dashed] node[draw,color=gray!50,minimum
        height=#1,minimum width=#2] (#4) at #5 {}; 
        \node[anchor=#3,inner sep=2pt] at (#4.#3)  {#6};
}

\node(client)at(1.35,0.7)[draw, align=left, rounded corners, text width=6.92cm, 
font=\tiny]{
\begin{tabular}{lc}
\textbf{SELECT} ?predicate ?object \textbf{WHERE} \{  & \\
 $dbpedia:Barack\_Obama$ ?predicate ?object \}  & ($tp_{1\_{CD1}}$) \\
  UNION \{ & \\
  ?subject $owl:sameAs$ \ $dbpedia:Barack\_Obama$ . &($tp_{2\_{CD1}}$) \\
  ?subject ?predicate ?object \} \} & ($tp_{3\_{CD1}}$)  \\
  \end{tabular}};

\node(client2)at(1.7,-1.3)[draw, align=left, rounded corners, text width=7.62cm, 
font=\tiny]{
\begin{tabular}{lc}
\textbf{SELECT} ?party ?page \textbf{WHERE} \{ & \\
$dbpedia:Barack\_Obama$ \ $dbpedia-owl:party$ ?party .  & ($tp_{1\_{CD2}}$) \\
?x $nytimes:topicPage$ ?page .  & ($tp_{2\_{CD2}}$) \\
?x $owl:sameAs$ $dbpedia:Barack\_Obama$ \}  & ($tp_{3\_{CD2}}$) \\
  \end{tabular}};

\node(client3)at(1.9,-3.6)[draw, align=left, rounded corners, text width=8.13cm, 
font=\tiny]{
\begin{tabular}{lc}
\textbf{SELECT} ?pres ?party ?page \textbf{WHERE} \{  & \\
?pres $rdf:type~$ \ $ dbpedia-owl:President$  .  & ($tp_{1\_{CD3}}$) \\
?pres $dbpedia-owl:nationality~$ \ $dbpedia:United\_States$ . & ($tp_{2\_{CD3}}$) \\
?pres $dbpedia-owl:party$ ?party . &($tp_{3\_{CD3}}$) \\
?x $nytimes:topicPage$ ?page . & ($tp_{4\_{CD3}}$) \\
?x $owl:sameAs$ ?pres  \} & ($tp_{5\_{CD3}}$)
  \end{tabular}};


\entity[A3,(7.7,0.6)] {$tp1\_{CD1}$};
\node(1)[text width=12em,minimum height=2em,font=\tiny] at(10.25,0.8){@DBPedia (Labels,};
\node(2)[text width=12em,minimum height=2em,font=\tiny] at(10.3,0.65){InstanceTypes,};
\node(3)[text width=12em,minimum height=2em,font=\tiny] at(10.3,0.5){Images, Articles,};
\node(4)[text width=12em,minimum height=2em,font=\tiny] at(10.25,0.35){InfoBox, NYTimes)};
\entity[B3,(7.7,-0.2)] {$tp2\_{CD1} \Join tp3\_{CD1}$};
\node(5)[text width=12em,minimum height=2em,font=\tiny] at(11.0,-0.2){@NYTimes};
\entity[C3,(7.7,-1)] {$tp1\_{CD2}$};
\node(6)[text width=12em,minimum height=2em,font=\tiny] at(10.3,-0.9){@DBPedia};
\node(36)[text width=12em,minimum height=2em,font=\tiny] at(10.3,-1.1){(InfoBox)};
\entity[C4,(7.7,-1.8)] {$tp3\_{CD2} \Join tp2\_{CD2}$};
\node(7)[text width=12em,minimum height=2em,font=\tiny] at(11.0,-1.8){@NYTimes};
\entity[C5,(7.7,-4.3)] {$tp4\_{CD3} \Join tp5\_{CD3}$};
\node(8)[text width=12em,minimum height=2em,font=\tiny] at(11.0,-4.3){@NYTimes};
\entity[C6,(7.7,-3.5)] {$tp2\_{CD3} \Join tp3\_{CD3}$};
\node(9)[text width=12em,minimum height=2em,font=\tiny] at(10.3,-2.6){@DBPedia};
\node(39)[text width=12em,minimum height=2em,font=\tiny] at(10.32,-2.8){(InstanceTypes)};
\entity[C7,(7.7,-2.7)] {$tp1\_{CD3}$};
\node(10)[text width=12em,minimum height=2em,font=\tiny] at(11,-3.4){@DBPedia};
\node(40)[text width=12em,minimum height=2em,font=\tiny] at(11,-3.6){(InfoBox)};

\node(11)[] at(7.8,1.5){\textbf{BGP$_{\textbf{1}}$}};
\node(12)[text width=20em,minimum height=2em,font=\tiny] at(9.5,1.1){\textbf{<QE$_{i}$,[11:24:19]-[11:24:32]>}};

 \draw[densely dashed] (C5) .. node[above] {} controls (6.3,-4.45) and (6.1,-2.55) ..  (C7); 
   \draw[densely dashed] (C5) .. node[above] {} controls (6.1,-4.45) and (6.05,-1.7) ..  (C4); 
   
\node(13)[text width=12em,minimum height=2em,font=\small] at(3.5,1.6){\textbf{CD1}};
\node(14)[text width=12em,minimum height=2em,font=\small] at(3.5,-0.5){\textbf{CD2}};
\node(15)[text width=12em,minimum height=2em,font=\small] at(3.5,-2.55){\textbf{CD3}};

\path[dashed] (A3) edge (B3)
 					(B3) edge (C3)
 					(C3) edge (C4)
 					(C7) edge (C6);
 \path[arrow,thick] (C6) edge (C5);

  \node(20)[] at(3.0,-4.8){};
  \node(21)[] at(4.4,-4.8){};
  \node(22)[] at(3.0,-5.1){};
  \node(23)[] at(4.4,-5.1){};
  \node(24)[text width=12em,minimum height=2em,font=\tiny] at(5.1,-4.95){\textbf{symhash}};
    \node(25)[text width=12em,minimum height=2em,font=\tiny] at(5.3,-4.65){\textbf{nlfo}};
   \path[arrow,thick] (20) edge (21);
   		              
   \path[densely dashed] (22) edge (23);
         
      
\end{tikzpicture}
 \caption{\feta's deduced $BGP_{1}$ for CD concurrent execution.}
	  \label{fig:feta_cd_output1}
	\end{figure}
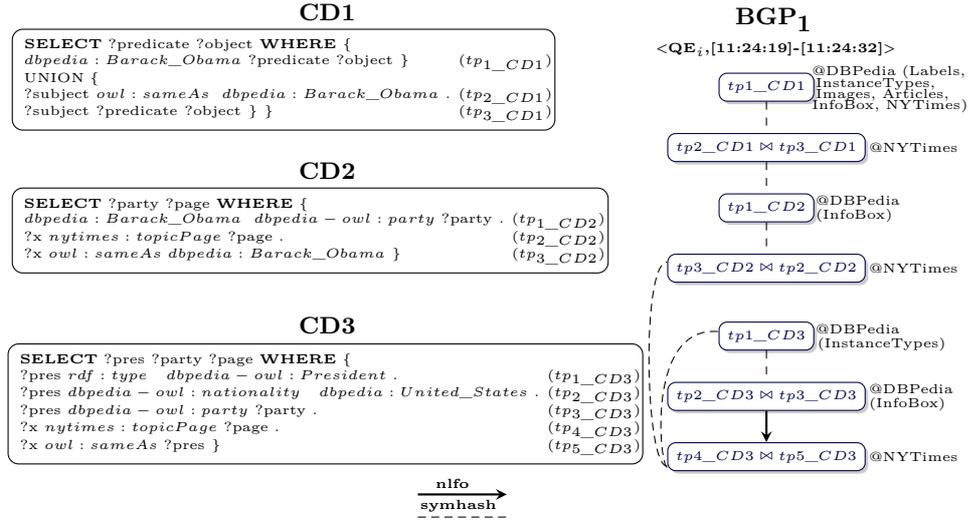

			\begin{figure}[ht!]
			\begin{tikzpicture}

			\tikzstyle{vecArrow} = [thick, decoration={markings,mark=at position
			   1 with {\arrow[semithick]{open triangle 60}}},
			   double distance=3pt, shorten >= 5.5pt,
			   preaction = {decorate},
			   postaction = {draw,line width=2pt, white,shorten >= 4.5pt}]
			\def\kbbox[#1,#2,#3,#4,#5]#6{
			        \draw[densely dashed] node[draw,color=gray!50,minimum
			        height=#1,minimum width=#2] (#4) at #5 {}; 
			        \node[anchor=#3,inner sep=2pt] at (#4.#3)  {#6};
			}
			     
			\node(client5)at(9.8,-0.6)[draw, align=left, rounded corners, text width=7.45cm, 
			font=\tiny]{
			\begin{tabular}{lc}
			\textbf{SELECT} ?film ?director ?genre \textbf{WHERE} \{  & \\
			?film \ $dbpedia-owl:director$ ?director .  & ($tp_{1\_{CD5}}$) \\
			?director $dbpedia-owl:nationality~$ \ $dbpedia:Italy$ . & ($tp_{2\_{CD5}}$)\\
			?x \ $owl:sameAs$ ?film . & ($tp_{3\_{CD5}}$) \\
			?x \ $linkedMDB:genre$ ?genre \} & ($tp_{4\_{CD5}}$)
			  \end{tabular}};

			\entity[C8,(15.55,-0.2)] {$tp_{2\_{CD5}}\Join tp_{1\_{CD5}}$};
			\node(30)[text width=12em,minimum height=2em,font=\tiny] at(18.85,-0.15){@DBPedia};
		    \node(40)[text width=12em,minimum height=2em,font=\tiny] at(18.9,-0.33){(InfoBox)};
			\entity[C9,(15.55,-1.2)] {$tp_{4\_{CD5}} \Join tp_{3\_{CD5}}$};
			\node(31)[text width=12em,minimum height=2em,font=\tiny] at(18.85,-1.2){@LMDB};
		
			
		\node(34)[] at(15.55,0.6){\textbf{BGP$_{\textbf{2}}$}};
		\node(35)[text width=20em,minimum height=2em,font=\tiny] at(17.2,0.25){\textbf{<QE$_{i}$,[11:24:24]-[11:24:30]>}};
			 
			\node(16)[text width=12em,minimum height=2em,font=\small] at(11.4,0.33){\textbf{CD5}};
			
			 \path[arrow,thick] (C8) edge (C9);
			 
			  \node(20)[] at(11.1,-1.7){};
			   \node(21)[] at(12.5,-1.7){};
			   \node(22)[] at(11.1,-2.0){};
			   \node(23)[] at(12.5,-2.0){};
			   \node(24)[text width=12em,minimum height=2em,font=\tiny] at(13.2,-1.85){\textbf{symhash}};
			     \node(25)[text width=12em,minimum height=2em,font=\tiny] at(13.4,-1.55){\textbf{nlfo}};
			   \path[arrow,thick] (20) edge (21);
			   		              
			   \path[densely dashed] (22) edge (23);

			\end{tikzpicture}
			 \caption{\feta's deduced $BGP_{2}$ for CD concurrent execution.}
				  \label{fig:feta_cd_output2}
				\end{figure}
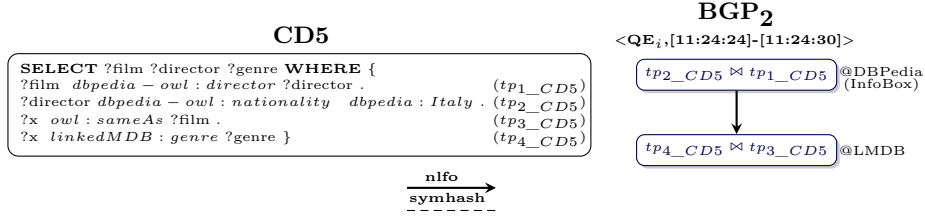

	\begin{figure}[ht!]
	\begin{tikzpicture}

	\tikzstyle{vecArrow} = [thick, decoration={markings,mark=at position
	   1 with {\arrow[semithick]{open triangle 60}}},
	   double distance=3pt, shorten >= 5.5pt,
	   preaction = {decorate},
	   postaction = {draw,line width=2pt, white,shorten >= 4.5pt}]
	\def\kbbox[#1,#2,#3,#4,#5]#6{
	        \draw[densely dashed] node[draw,color=gray!50,minimum
	        height=#1,minimum width=#2] (#4) at #5 {}; 
	        \node[anchor=#3,inner sep=2pt] at (#4.#3)  {#6};
	}

	\node(client6)at(0,-0.6)[draw, align=left, rounded corners, text width=8cm, 
	font=\tiny]{
	\begin{tabular}{lc}
	\textbf{SELECT} ?name ?location \textbf{WHERE} \{  & \\
	?artist \ $foaf:name$ ?name .  & ($tp_{1\_{CD6}}$) \\
	?artist \ $foaf:based\_near$ ?location . & ($tp_{2\_{CD6}}$)  \\
	?location \ $geonames:parentFeature$ ?germany . &($tp_{3\_{CD6}}$) \\
	?germany \ $geonames:name$   "Federal Republic of Germany" \} & ($tp_{4\_{CD6}}$) 
	  \end{tabular}};

	   		      \entity[C12,(5.55,0)] {$tp_{4\_{CD6}} \Join tp_{3\_{CD6}}$};
	   		      \node(40)[text width=12em,minimum height=2em,font=\tiny] at(8.82,0){@Geonames};
	   		      \entity[C13,(5.55,-0.7)] {$tp_{2\_{CD6}}$};
	   		      \node(41)[text width=12em,minimum height=2em,font=\tiny] at(8.15,-0.6){@Jamendo, @SWDF};
	   		      \node(47)[text width=12em,minimum height=2em,font=\tiny] at(8.15,-0.75){@LMDB};
	   		      \entity[C14,(5.55,-1.4)] {$tp_{1\_{CD6}}$};
	   		      \node(42)[text width=13em,minimum height=2em,font=\tiny] at(8.35,-1.2){@DBPedia, };
	   		      \node(43)[text width=12em,minimum height=2em,font=\tiny] at(8.2,-1.4){(InfoBox, Person)};
	   		      \node(45)[text width=12em,minimum height=2em,font=\tiny] at(8.2,-1.6){@Jamendo, @SWDF};

	   \path[densely dashed]	(C12) edge (C13)
	   		               		(C13) edge (C14);
	 
	\node(16)[text width=12em,minimum height=2em,font=\small] at(1.3,0.33){\textbf{CD6}};

		\node(34)[] at(5.55,0.8){\textbf{BGP$_{\textbf{3}}$}};
		\node(35)[text width=20em,minimum height=2em,font=\tiny] at(7.2,0.45){\textbf{<QE$_{i}$,[11:24:25]-[11:24:33]>}};

	  \node(20)[] at(1.1,-2){};
	  \node(21)[] at(2.5,-2){};
	  \node(22)[] at(1.1,-2.3){};
	  \node(23)[] at(2.5,-2.3){};
	  \node(24)[text width=12em,minimum height=2em,font=\tiny] at(3.2,-2.15){\textbf{symhash}};
	  \node(25)[text width=12em,minimum height=2em,font=\tiny] at(3.4,-1.85){\textbf{nlfo}};
	   \path[arrow,thick] (20) edge (21);
	   		              
	   \path[densely dashed] (22) edge (23);

	\end{tikzpicture}
	 \caption{\feta's deduced $BGP_{3}$ for CD concurrent execution.}
		  \label{fig:feta_cd_output3}
		\end{figure}
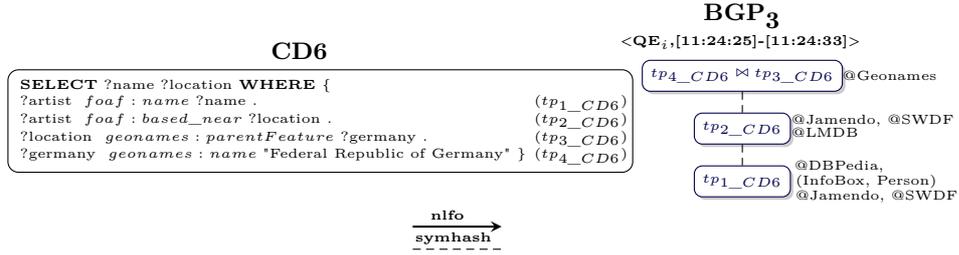
			
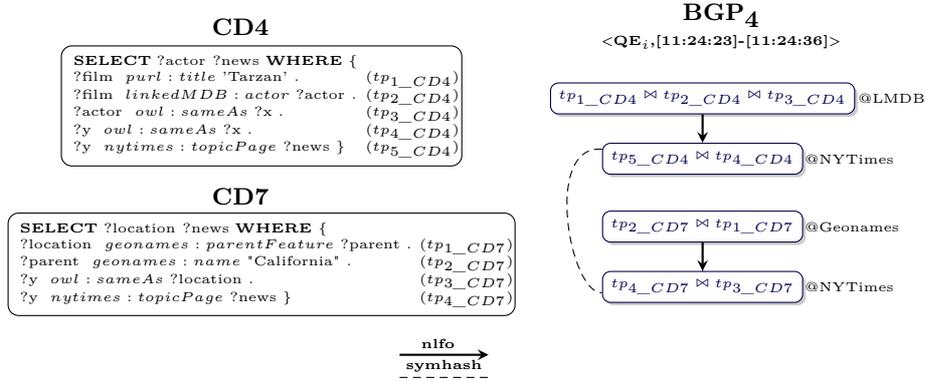
\begin{figure}[ht!]
\begin{tikzpicture}
\tikzstyle{vecArrow} = [thick, decoration={markings,mark=at position
   1 with {\arrow[semithick]{open triangle 60}}},
   double distance=3pt, shorten >= 5.5pt,
   preaction = {decorate},
   postaction = {draw,line width=2pt, white,shorten >= 4.5pt}]
\def\kbbox[#1,#2,#3,#4,#5]#6{
        \draw[densely dashed] node[draw,color=gray!50,minimum
        height=#1,minimum width=#2] (#4) at #5 {}; 
        \node[anchor=#3,inner sep=2pt] at (#4.#3)  {#6};
}

 \node(client4)at(1.2,-0.5)[draw, align=left, rounded corners, text width=5.05cm, 
 font=\tiny]{
 \begin{tabular}{lc}
 \textbf{SELECT} ?actor ?news \textbf{WHERE} \{ & \\
 ?film  \ $purl:title$  'Tarzan' .  & ($tp_{1\_{CD4}}$) \\
 ?film \ $linkedMDB:actor$  ?actor .  & ($tp_{2\_{CD4}}$) \\
 ?actor \ $owl:sameAs$  ?x .  & ($tp_{3\_{CD4}}$) \\
 ?y \ $owl:sameAs$  ?x .  & ($tp_{4\_{CD4}}$) \\
 ?y  \ $nytimes:topicPage$  ?news \}  & ($tp_{5\_{CD4}}$) \\
   \end{tabular}};

  \node(client7)at(1.2,-2.6)[draw, align=left, rounded corners, text width=6.45cm, 
  font=\tiny]{
  \begin{tabular}{lc}
  \textbf{SELECT} ?location ?news \textbf{WHERE} \{  & \\
  ?location \ $geonames:parentFeature$ ?parent .  & ($tp_{1\_{CD7}}$)  \\
  ?parent \ $geonames:name$ "California" . & ($tp_{2\_{CD7}}$) \\
  ?y \ $owl:sameAs$ ?location . & ($tp_{3\_{CD7}}$) \\
  ?y \ $nytimes:topicPage$ ?news \} & ($tp_{4\_{CD7}}$)
    \end{tabular}}; 

\entity[C8,(7.,-0.4)] {$tp_{1\_{CD4}} \Join tp_{2\_{CD4}} \Join tp_{3\_{CD4}}$};
\node(30)[text width=12em,minimum height=2em,font=\tiny] at(11.,-0.4){@LMDB};
\entity[C9,(7.,-1.2)] {$tp_{5\_{CD4}} \Join tp_{4\_{CD4}}$};
\node(31)[text width=12em,minimum height=2em,font=\tiny] at(10.3,-1.2){@NYTimes};
 \path[arrow,thick] (C8) edge (C9);

\entity[C15,(7.,-2.1)] {$tp_{2\_{CD7}}\Join  tp_{1\_{CD7}}$};
\node(40)[text width=12em,minimum height=2em,font=\tiny] at(10.3,-2.1){@Geonames};
\entity[C16,(7.,-2.9)] {$tp_{4\_{CD7}} \Join tp_{3\_{CD7}}$};
\node(41)[text width=12em,minimum height=2em,font=\tiny] at(10.3,-2.9){@NYTimes};

\node(38)[] at(7.25,0.7){\textbf{BGP$_{\textbf{4}}$}};
\node(39)[text width=20em,minimum height=2em,font=\tiny] at(8.92,0.35){\textbf{<QE$_{i}$,[11:24:23]-[11:24:36]>}};

 \draw[densely dashed] (C9) .. node[above] {} controls (4.9,-1.0) and (5.25,-3.0) ..  (C16); 
 
\node(18)[text width=12em,minimum height=2em,font=\small] at(2.5,0.55){\textbf{CD4}};
\node(19)[text width=12em,minimum height=2em,font=\small] at(2.5,-1.7){\textbf{CD7}};

\path[arrow, thick] (C15) edge (C16);
  
  \node(20)[] at(2.9,-3.8){};
  \node(21)[] at(4.3,-3.8){};
  \node(22)[] at(2.9,-4.1){};
  \node(23)[] at(4.3,-4.1){};
  \node(24)[text width=12em,minimum height=2em,font=\tiny] at(5.05,-3.95){\textbf{symhash}};
    \node(25)[text width=12em,minimum height=2em,font=\tiny] at(5.25,-3.65){\textbf{nlfo}};
   \path[arrow,thick] (20) edge (21);
 
   	 \path[densely dashed]	(C15) edge (C16);
   		               	    
   \path[densely dashed] (22) edge (23);

\end{tikzpicture}
	 \caption{\feta's deduced $BGP_4$ for CD concurrent execution.}
	  \label{fig:feta_cd_output4}
	\end{figure}

%% file: relatedWork.tex
\section{Related Work}
\label{sec:related_work}

Federated query tracking is related to web
tracking~\cite{web_tracking_12}. In web tracking, a first-party
website authorizes a third party to learn about its users. Analogously, \feta plays the role of the third party.
However, logs collected in federated query tracking is the result of
the execution of a physical plan in distributed query processing
compared to a more simple web navigation flow in web tracking. 

Extracting information from raw logs is traditionally a data mining 
process~\cite{han2011data_mining} involving the following steps:
\begin{inparaenum}[(i)] 
\item \emph{data selection} identifies the
  target dataset and relevant attributes that will be used to derive
  new information; 
\item \emph{data cleaning} removes noise and
  outliers, transforms field values to common units, generates new
  fields and finally brings the data into the structured data schema
  that is used for storage, e.g., relational databases, XML;
\item \emph{data mining} applies data analysis and discovery algorithms
  based on machine learning, pattern recognition, statistics and other
  methods; 
\item finally \emph{evaluation} presents the new knowledge
  in a form that will be also understandable from the end user, e.g.,
  through visualization.

\end{inparaenum}

\feta can be located at the data mining step because it transforms raw logs
into a sequence of sets of BGPs. In experiments, we presented one
extraction for a period of time. Repeating extractions will generate a
sequence of extracted BGPs than can be used for association rules
mining of frequent pattern detection.

%% file: conclusion.tex
\section{Conclusions and future work}
\label{sec:conclusion}

Federated query tracking allows data providers to access secondary data in Linked Data federation. We proposed \feta, a federated
query tracking approach that extracts original federated Basic Graph Patterns
from a shared log maintained by data providers. \feta links and
unlinks variables present in different subqueries of the federated log
by applying a set of heuristics we presented in this paper.

Even in a worst case scenario with Anapsid, \feta extracts BGPs that contain
original BGPs of federated join queries. Extracted BGPs with endpoints'
information give valuable information to data providers about which triples are joined, when and by whom.

We think \feta opens several interesting perspectives. First,
heuristics can be improved in many ways by better using semantics of
predicates and answers. Second, we conducted experiments on one slice of federated
log. Repeating BGPs extractions on successive slices allows to apply
traditional data mining techniques such as extracting frequents
patterns. Third, we limited experiments to Anapsid. Extending to FedX
query engine~\cite{fedex-fluid-schwarte-11} will challenge proposed
heuristics because FedX physical operators produce slightly
different query traces.